# Post-field ionization of Si clusters in atom probe tomography: A joint theoretical and experimental study

Ramya Cuduvally*[1,2], Richard J. H. Morris[2], Giel Oosterbos[1], Piero Ferrari[1], Claudia Fleischmann[1,2], Richard G. Forbes[3], Wilfried Vandervorst[1,2]

[1.] *Quantum Solid-State Physics, KU Leuven, Celestijnenlaan 200D, 3001 Leuven, Belgium.*

[2.] *IMEC, Kapeldreef 75, 3001, Leuven, Belgium.*

[3.] *Advanced Technology Institute & Department of Electrical and Electronic Engineering,*
*University of Surrey, Guildford, Surrey GU2 7XH, UK.*

*Current address: Canadian Centre for Electron Microscopy, McMaster University, 1280 Main St West, Hamilton, Ontario, L8S 4M1, Canada*

**Abstract:** A major challenge for Atom Probe Tomography (APT) quantification is the inability to decouple ions which possess the same mass-charge (m/n) ratio but a different mass. For example, $^{75}As^+$ and $^{75}As_2^{2+}$ at ~75 Da or $^{14}N^+$ and $^{28}Si^{2+}$ at ~14 Da, cannot be differentiated without the additional knowledge of their kinetic energy or a significant improvement of the mass resolving power. Such mass peak overlaps lead to ambiguities in peak assignment, resulting in compositional uncertainty and an incorrect labelling of the atoms in a reconstructed volume. In the absence of a practical technology for measuring the kinetic energy of the field-evaporated ions, we propose and then explore the applicability of a post-experimental analytical approach to resolve this problem based on the fundamental process that governs the production of multiply charged molecular ions/clusters in APT, i.e., Post-Field Ionization (PFI). The ability to predict the PFI behaviour of molecular ions as a function of operating conditions could offer the first step towards resolving peak overlap and minimizing compositional uncertainty. We explore this possibility by comparing the field dependence of the charge-state-ratio for Si clusters ($Si_2$, $Si_3$ and $Si_4$) with theoretical predictions using the widely accepted Kingham PFI theory. We then discuss the model parameters that may affect the quality of the fit and the possible ways in which the PFI of molecular ions in APT can be better understood. Finally, we test the transferability of the proposed approach to different material systems and outline ways forward for achieving more reliable results.

# 1. Introduction

Atom Probe Tomography (APT) is an indispensable pathway for the investigation of nanoscale structure-property relationships in various scientific disciplines including metallurgy [1], electronics [2–5] geology [6] and biology [7]. Quantitative compositional information in APT is contained in the mass spectrum, which is constructed from the time-of-flight of the ions and using the law of energy conservation. While several factors may influence the accuracy of the determined composition, the most important factor is the ability to unambiguously assign the correct chemical identity to each peak observed in the mass spectrum based on its mass/charge (m/n) ratio. Therefore, ions which possess a different mass but the same m/n ratio (i.e., overlapping peaks) present a problem for APT quantification and, subsequently, the reconstruction. For example, it is not possible to differentiate between ions such as $^{14}N^+$ and $^{28}Si^{2+}$ at ~14 Da or molecular ions/clusters of the same species, e.g., $^{75}As_2^+$ and $^{75}As_4^{2+}$ at ~150 Da. Such mass peak overlaps will lead to ambiguities in peak assignment, and ultimately result in a compositional uncertainty.

For the case of the complete overlap between singly and multiply charged clusters, e.g., $^{75}As_2^+$ and $^{75}As_4^{2+}$ at ~150 Da, potentially, one could differentiate them discreetly [8] based on their kinetic energy. Although research is currently focused on enabling kinetic energy discrimination for APT [8,9], no such technology has yet been practically implemented for commercial instruments since they are still in their initial stage of development. In some cases where the overlapping species has more than one stable isotope, it may be possible to use isotope-constrained methods [10] to (collectively [8]) discriminate them. However, such techniques are essentially inapplicable to mono-isotopic species such as As and P that are important elements in many nano-electronic devices for which APT is extensively used in fabrication process development and control. With the advent of III-V electronics, a plethora of compound materials including GaAs, InGaAs, InP, InGaAsP are now found ubiquitously in optoelectronics, photonics, photovoltaics and high-speed analog/RF devices [11–14]. Furthermore, As and P are commonly used as the n-type dopants in the more conventional Si or SiGe based electronic devices [15,16]. The problems associated with As and P are enhanced further as both these elements have a preferential tendency to field-evaporate as clusters [17,18], thus increasing the probability for numerous peak overlaps and consequently a huge composition quantification uncertainty.

An alternative solution for improving quantification accuracy would be through accurately predicting the level of peak overlap between cluster ions based on the analysis conditions. In this work, we outline a novel approach to achieve this using the Post-Field Ionization (PFI) theory developed for monoatomic ions by Kingham [19] and adapting for clusters. The cluster ionization energies, which are unknown, were computed using Density Functional Theory (DFT). Subsequently, the PFI probability for the various charge-states of the clusters was estimated as a function of the electric field. The validity of these predictions was then examined against experimental APT data of Si operating in laser-assisted mode. Si was chosen because it exhibits extensive clustering (up to $Si_7$) for certain operating conditions [20] and has three stable isotopes. Hence, the peak overlap between the various Si clusters, e.g., $^{28}Si^+$ and $^{28}Si_2^{2+}$ at ~28 Da and $^{28}Si_2^+$ and $^{28}Si_4^{2+}$ at 56 Da, are resolvable using the conventional isotopic peak deconvolution approach. This enabled a quantitative evaluation of our proposed theoretical approach as a stand-alone technique for resolving peak overlap. Through this joint theoretical and experimental study, we show that the field-dependence of the charge-state of Si cluster ions in APT is qualitatively consistent with the predictions of Kingham's PFI theory. We also discuss the reasons for the rather poor quantitative agreement observed between theory and experiment and further demonstrate how improvements can be made to this model based on physically justifiable arguments. Finally, we will assess the transferability of the proposed approach by applying it to resolve overlaps between As clusters that are observed in the APT InGaAs [17].

## 1.1 Theory

In this section, the theoretical principles governing the PFI model proposed by Kingham [19] will be outlined. We will also highlight some of the criticism that the Kingham model received in the later years and the associated implications. Since not all derivations/assumptions were explicitly stated in Kingham's paper [19], reproducing the original calculations is non-trivial. A comprehensive description of the theoretical framework is included in the supplementary section.

The presence of multiply charged species in APT finds its origin in the processes of field-evaporation and PFI. The latter is a process by which a field-evaporated ion (that is being accelerated away from the emitter) loses an additional electron by quantum tunneling from the ion into an empty electronic state of the emitter, or, into free space. An ion can even be ionized multiple times after field-evaporation. The PFI probability is determined by (and increases with) the electric field strength at the apex. Furthermore, the material properties (i.e., the ionization energy and surface work function) also influence the electron tunneling rate and subsequently the PFI probability. The first successful PFI investigations were related to the field evaporation of rhodium. In an experimental investigation, Ernst [21] showed that $Rh^{2+}$ was formed by PFI with the excellent agreement between Kingham's PFI model and experimental Rh data [19]. Later, Ernst and Jentsch [22] showed that the experiments [20] were consistent with a tunnelling mechanism as described above. Further work by Haydock and Kingham [23–25] confirmed these theoretical results, with Kingham [19] subsequently carrying out PFI calculations for many metallic ion species. Following this, Kellogg [26] experimentally investigated the PFI behavior for six other elements and found a satisfactory agreement between Kingham's model and four of these elements (details in section 1.2). Thus, the general validity of PFI through the Kingham model was established. The Kingham model describes the electric field dependence of the Charge-State-Ratio (CSR), i.e., $X^{i+}/\sum_i X^{i+}$, ($i$ = 1, 2…), of any given constituent species (X) in an APT experiment. Since there is currently no easy way to measure the electric field at the apex of a specimen, the Kingham model is typically used to estimate the magnitude of the electric field based on the CSR, which is readily extracted from the mass spectrum.

In the Kingham model [19], a single parameter fit was used to match Rh experimental data from Ernst [21], in which the field strength values were determined by a calibration method described in [27] with an absolute error of at least 15%. The fitting parameter used by Kingham was the effective nuclear potential, Z, that is seen by the tunneling electron. As previously mentioned, the validity of the model was verified for several other monoatomic ions experimentally. On the other hand, the PFI behavior of molecular/cluster ions have received much less attention thus far. As far as the authors are aware, only three studies [28–30] have been reported in the literature pertaining to the post-ionization behavior and observed charge-state of cluster ions in APT and some of these findings will be discussed in section 1.2. However, the applicability of PFI theory and the validity of Kingham's PFI model have not been experimentally examined thus far for any molecule/cluster. Such a theoretical and experimental comparison is important because the knowledge of cluster PFI behavior as a function of apex field offers the first step towards tackling quantification inaccuracies resulting from peak overlap. For example, in the APT analysis of InGaAs, resolving the overlap between $As_2^+$ and $As_4^{2+}$ (150 Da) using the PFI theory would involve the following steps.

A. Estimate the field strength using the CSR of a monoatomic species in conjunction with its Kingham curve or, alternatively, another field calibration technique [26,27].
B. Predict the ratio of $As_4^{2+}/ (As_4^+ + As_4^{2+})$ at this field using a theoretical PFI model that is applicable to molecules.
C. Based on (B) and from the number of counts in the $As_4^+$ peak (which is assumed to not suffer from overlap with $As_8^{2+}$), calculate the amount of $As_4^{2+}$ present in the peak at 150 Da.
D. Subtract the value found in (C) from the total counts measured at 150 Da to determine the number of $As_2^+$ ions.

The most important requirement for the procedure described above is B, i.e., the availability of a reliable PFI model for clusters. Following the approach used by Gault *et al.* [29] and as a first approximation, we extended Kingham's model to clusters by simply replacing the physical quantities of the monoatomic ions with their respective molecular equivalents (this primarily concerns the ionization energies). We then tested the applicability/accuracy of this model using experimental data of CSR versus electric field for Si clusters ($Si_2 - Si_4$). Thus, the objectives of this study were fourfold:

1) To explore the CSR – field relationship for Si clusters ($Si_2$, $Si_3$ and $Si_4$) experimentally and compare this with the predictions of PFI theory. This is the first attempt to model the PFI behavior of clusters based on experimental data.
2) To identify the model parameters that may need modification when considering the PFI of clusters and evaluate the impact of varying each of these parameters on the PFI curves.
3) To establish whether the predictions are accurate enough to be used for resolving peak overlap between clusters of mono-isotopic species.
4) Build an understanding of the PFI of molecules in APT and outline the ways forward for improving the agreement of the molecular PFI model with experimental data.

## 1.2 A critical look at the Kingham model - a review of literature

The biggest challenge encountered in testing the validity/quantitative accuracy of the analytical PFI model lies in the lack of a method to directly measure the electric field at the specimen apex. Therefore, indirect schemes need to be employed to estimate a value for the field and this can be subject to significant errors. Thus, any PFI model will only be as good as the apex field estimation method used. Kingham's work [19] was followed up with experimental investigations by Kellogg [26], wherein the author compared the experimental charge-state evolution of Ni, Mo, Rh, W, Re, Ir and Pt with the predictions of Kingham's PFI model. To determine the field strength, F, in the experiment, Kellogg used the following calibration formula:

$$F = F_0(V/V_0) \qquad \text{E1}$$

Here, $F_0$ is evaporation field threshold at 0 K, V is the applied voltage and $V_0$ is the threshold (DC) voltage for field-evaporation at the base temperature (and in the absence of laser illumination). Note that this field determination method was different to that of Ernst for Rh [21], which was based on a one-time calibration of the voltage as a function of the electric field by Müller and Young [27], which in turn was based on a corrected version of the original Fowler-Nordheim theory of field-electron emission. The accuracy of the field values determined by Kellogg and Ernst was no less than 25% and 15%, respectively [19,26]. Although Kellogg observed some deviations from the Kingham model at low field for W, Re and Ir, in general, the field value at which one charge-state becomes more abundant than the other was found to be in reasonable agreement with that predicted by the model. For Mo and Pt however, Kellogg found that although the experimental results confirmed the occurrence of PFI, the agreement with Kingham's model was poor. In the same paper, Kellogg also experimentally investigated the PFI of Si. However, the $F_0$ value used for field calibration was not taken from the literature but adjusted to yield the best fit to the Kingham model. The value chosen was 35 V/nm. Therefore, for Si, only a qualitative agreement with PFI trends could be established.

This qualitative agreement of the PFI behavior of Si with the Kingham model was also confirmed by Kumar *et al.* in a separate study [31]. However, they used an $F_0$ value of 27.5 V/nm to obtain the best fit to the Kingham curves. Note that significant variations in $F_0$ can be a result of surface contaminants [26] and/or local differences in the binding energy between the atom and the surface [32]. More recently, Schreiber *et al.* [28] explored the post-ionization behavior of Fe and O in the APT characterization of $Fe_3O_4$. However, their objective was to investigate whether PFI was the driving factor for the formation of multiply charged species and if the fraction of multiply charged species increased with field. Interestingly, they found that the Fe CSR value was not exclusively dictated by field strength but also showed a dependence on the laser pulse energy used. Conversely, for O, they found that the fraction of $O_2$ clusters was purely field dependent and thus the ratio $O^+/ (O^+ + O_2^+)$ was a more reliable indicator of the field compared to the Fe CSR. Again, no quantitative comparisons were made with the Kingham model. In another study, Gault *et al.*[29] extended the Kingham model to $N_2$ in order to determine how likely the formation of $N_2^{2+}$ was at the typical values of field strength used in their experiments. They did so by replacing the atomic IEs with those for the molecular ($N_2$). More recently, in their quest for a statistically reliable CSR to indicate the field, the same approach was applied to the $TiD_2$ molecule by Chang *et al.* [30]. As mentioned previously, this is also the method we will use for Si clusters in this work. However, in their calculation of the Kingham curves, the authors ignored the region close to the tip surface (thus rendering the crossing surface/critical distance irrelevant). We believe that it is important to include the region close to the tip surface in computing the electron tunneling probability since this region is expected to contribute significantly to this probability. This, and the lack of a standardized software/script for the computation of these curves, motivated our own calculations of the Kingham curves and this is described in detail in the supplementary section.

## 2. Workflow and methods

### 2.1 Workflow

#### 2.1.1 Experimental

The details of the experimental data acquisition and treatment are listed below.

1) Acquisition of experimental data from a standard CAMECA Si PSM specimen [33] in laser mode, which allows the evaporation/detection of Si clusters in the mass spectrum.

2) Application of isotopic peak deconvolution to resolve any mass spectrum peak overlaps between Si and $Si_2^{2+}$, $Si_2^+$ and $Si_4^{2+}$, and $Si_3^+$ and $Si_6^{2+}$.

3) Calculation of the CSR of Si clusters, i.e., $Si_2^{n+}/(Si_2^+ + Si_2^{2+})$, $Si_3^{n+}/(Si_3^+ + Si_3^{2+})$ and $Si_4^{n+}/(Si_4^+ + Si_4^{2+})$, where n = 1, 2.

4) Estimation of the electric field strength for each operating condition using the Si CSR, i.e., $Si^{n+}/(Si^+ + Si^{2+})$, where n = 1 or 2, in conjunction with the Kingham curve of monoatomic Si.

5) Generation of the experimental cluster PFI curves, i.e., the variation in CSR of Si clusters ($Si_2$, $Si_3$ and $Si_4$) as a function of field strength.

### 2.1.2 Theoretical

In the framework proposed by Kingham [19], the PFI probability of an ion depends on its ionization energy (I), its atomic mass (m), the surface work function ($\phi$), and the principal quantum number of the highest occupied orbital of the atom ($m_q$). To extend Kingham's model to clusters, the following steps/assumptions were employed.

1) Since experimental/theoretical values of the higher order IEs of Si clusters were unavailable in the literature, the Vertical IEs of Si clusters were computed using DFT (details in sections 2.2, 3.1, supplementary section and [34]).

2) Using the computed values of IE and the equations for the ionization rate constant (R), ion velocity (u), and post-field-ionization probability ($P_t$) as given in section 1 of the supplementary section, the PFI curves for the Si clusters were computed.

3) The computed PFI curves were then compared with experimental curves (obtained in step 5, section 2.1.1).

4) Finally, the parameters that could potentially affect how well the model agrees with the experimental data are discussed.

### 2.2 Definitions: Adiabatic and Vertical Ionization Energy

The loss of an electron from a molecule may be accompanied by changes in its geometry since the ground state geometrical configuration of the ion typically differs from that of the neutral. Thus, two variants of the IE can be defined: Adiabatic Ionization Energy (AIE) and Vertical Ionization Energy (VIE) [35]. The AIE corresponds to the *minimum* energy required to remove an electron from a neutral molecule and is defined as the energy difference between the vibrational ground state of the neutral and the ion. An adiabatic ionization involves a rearrangement of the molecular nuclei, which is a much slower process than any electronic transition. The Vertical Ionization Energy (VIE), in contrast, corresponds to the energy difference between the neutral and the ion without any change in the geometrical configuration of the constituent nuclei. Since the Franck-Condon approximation [36] dictates that a vertical electronic transition is more favorable than an adiabatic transition, for the ionization energies and the PFI curves of the Si clusters reported here, we chose to compute and use the VIEs.

## 2.3 Methods

The experimental data were acquired on the Cameca LEAP 5000 XR from Cameca Pre-Sharpened Microtip (PSM) Array Coupons [33] consisting of Sb doped Si specimens. UV laser pulse energies over the range 90 – 165 pJ were used. Three different specimens from two different PSM array coupons were measured, and the base temperature was maintained at 25, 27 or 50 K. For each operating condition, ~1 million ions were collected and the LPE was varied such that Si clusters (up to $Si_4$ or $Si_5$) were visible in the mass spectrum. The operating conditions for the three different specimens (referred to as experiment 1, 2 and 3) are listed below. The objective of these experiments was to estimate the CSR of the Si clusters ($Si_2$, $Si_3$ and $Si_4$) as a function of the operating conditions. Subsequently, by determining the Si-CSR value at each of these operating conditions and using the Si Kingham curve, the value of the electric field was estimated at each operating condition. Thus, experimental PFI curves for the Si clusters could be generated.

*Experiment 1:*
(a) Constant LPE of 100 pJ and varying detection rate in steps from 0.5 to 3.5 %.
(b) Constant field (floating detection rate) and varying LPE in steps from 100 to 140 pJ. The minimum and maximum set-point values of detection rate were 0.5 and 17 %.
*Experiment 2:* Constant detection rate of 1 % and varying LPE in steps from 105 to 165 pJ.
*Experiment 3:* Constant detection rate of 0.5 % and varying LPE in steps from 110 to 125 pJ.

The atom probe data was subsequently analyzed using the commercially available CAMECA IVAS 3.8.4 and/or custom written MATLAB codes. The isotope-constrained peak deconvolution algorithm was applied to the experimental data using the built-in peak decomposition functionality within IVAS prior to calculating the CSRs of Si clusters. DFT calculations were performed with the ORCA 4.1.1 software package [37], to determine the Si cluster ionization energies. The range-separated hybrid ωB97X-D3 exchange-correlation functional [38], which includes a dispersion correction and the large def2-TZVP basis-set [39] were used. Prior to selecting the functional, several other functionals, namely, PBE, TPSS, B3LYP and the CAM-B3LYP-D3 were used, in addition to the ωB97X-D3, to compute the first Adiabatic Ionization Energy (AIE) and bond lengths of Si clusters (1- 7). These results were then compared with experimental data reported in the literature [40]. The details of these calculations along with the atomic coordinates of the energy optimized Si clusters are given in the supplementary section and previously reported by Oosterbos [34]. It was found that the ωB97X-D3 functional yielded the best match with experiment data and hence was chosen for the computation of higher order Vertical Ionization Energies (VIEs). The spin state corresponding to the most stable geometry for each cluster was determined and used here as previously reported by Oosterbos [34]. Finally, to ensure that the computed geometry and spin of the clusters corresponded to the true minima in the potential energy curve, the second derivative test was applied.

## 3. Results

### 3.1 Theoretical

The VIEs for the Si clusters calculated by DFT are given in Table 1. The details of the DFT calculations can be found in the supplementary section and [34]. For comparison, the IEs of Si [40,41] are also given. The primary interest for this study is the second IE of the clusters and these values are found to be smaller than the second IE of monoatomic Si (Table 1).

| Cluster size | 1st IE (eV) | 2nd IE (eV) | 3rd IE (eV) |
|---|---|---|---|
| 1 | 8.15 | **16.35** | 33.49 |
| 2 | 7.93 | **15.82** | 20.28 |
| 3 | 8.22 | **14.40** | 20.18 |
| 4 | 8.29 | **13.85** | 19.49 |

Table 1: VIEs of Si clusters ($Si_2$ -$Si_4$) computed with DFT using the ωB97X-D3 functional and the def2-TZVP basis-set. Si IEs are given in row 1 for comparison with clusters.

The values of VIE determined were subsequently used in the computation of the PFI curves for Si clusters, which are shown in Fig. 1. The other parameters, i.e., Z, $m_q$ and ϕ, used in the calculations were assumed to be identical to those of Si. These parameters will be re-visited in the following section. For each species, the electric field value corresponding to the (50 % CSR) crossover point is indicated by labels 1, 2, 3 and 4, which correspond to 19.6, 18, 14.4 and 13 V/nm, for Si, $Si_2$, $Si_3$ and $Si_4$ respectively. We will refer to this electric field value as the $F^{50}$ value from now on.

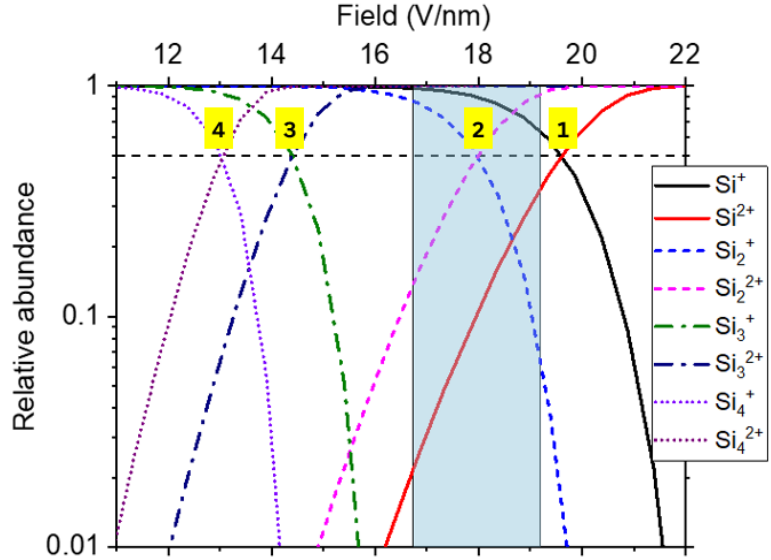

Fig. 1: PFI curves of Si and its clusters computed with an effective nuclear potential given by $Z = n + 1 + 4.5/z_0$, a principal quantum number, $m_q = 3$ and a surface work function, ϕ = 4.9 eV. Note that the blue region indicates the range of electric field within which the experimental data lie. The yellow labels 1, 2, 3 and 4 indicate the $F^{50}$ for Si, $Si_2$, $Si_3$ and $Si_4$ respectively.

## 3.2 Experimental

An example of a Si mass spectrum is shown in Fig. 2 and is representative of the mass spectra obtained for all the different operating conditions explored in this study. Although Si clusters up to $Si_5$ were detected, we limit our discussion to $Si_2$, $Si_3$ and $Si_4$ due to insufficient statistics of clusters larger than these. As expected, there were significant peak overlaps in the mass spectra, which were readily resolved using isotopic peak deconvolution. For example, when using a LPE of 100 pJ and a set point detection rate of 1% (Experiment 1), the $Si_2^{2+}$-CSR before and after applying the peak deconvolution algorithm was found to be 0.048 and 0.543. This is a significant deviation which would result in an incorrect conclusion about the accuracy of the theoretical predictions if not taken in account.

To visualize the trend in the experimental data and to account for the known tip asymmetry (and thus local differences in electric field strength), the dataset for each operating condition was divided into two regions, as shown in Fig. 3. The apex electric field-strength is expected to be lower on the laser side due to the larger local radius of curvature, which is reflected in the density plot (Fig. 3(a)) and the Si-CSR distribution (Fig. 3(b)).

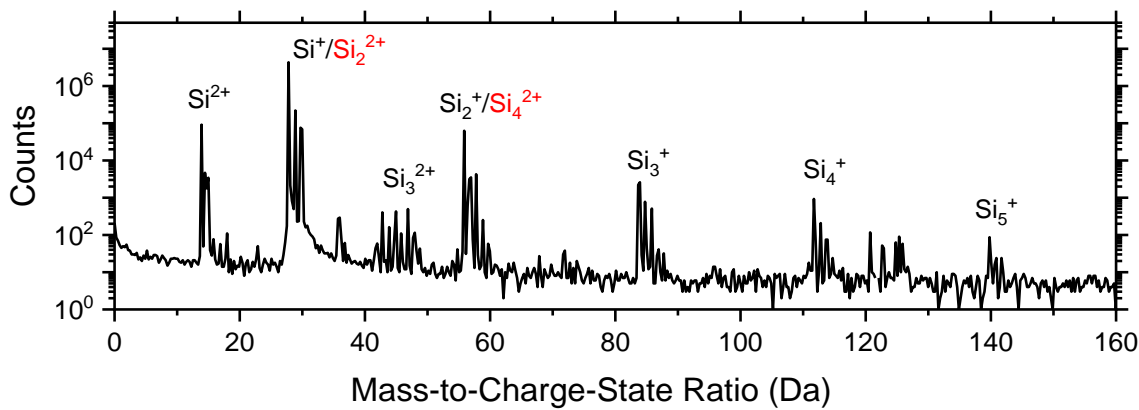

Fig. 2: Mass spectrum of Si obtained on the LEAP 5000 XR at an applied laser pulse energy of 125 pJ, $Si^{2+}/(Si^+ + Si^{2+}) = 0.02$, base temperature of 25 K and set point detection rate of 1 %.

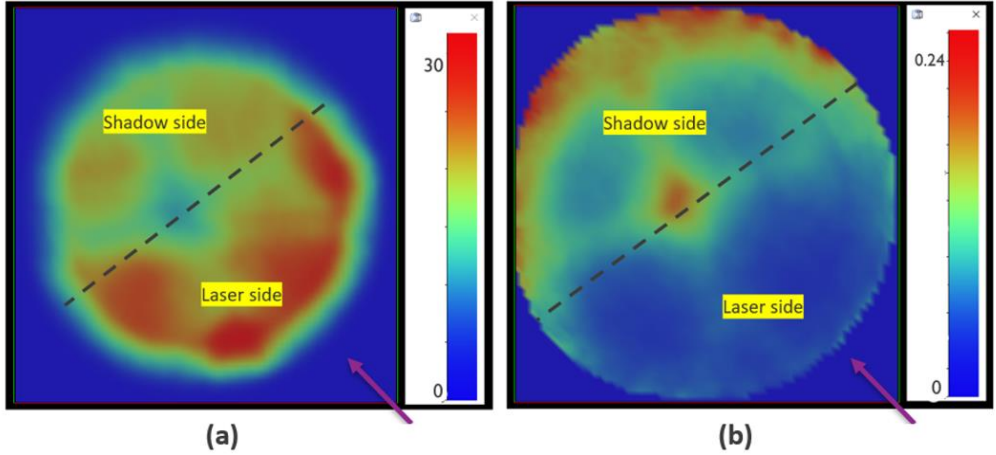

Fig. 3: An example of division of the dataset into two regions for data analysis. (a) 2D density plot of all ions (b) 2D concentration plot of $Si^{2+}$ CSR (i.e., $Si^{2+}/(Si^{2+} + Si^{+})$) at an LPE of 110 pJ and detection rate of 0.5 %. The laser direction is indicated by the violet arrow. Note that these plots are only representative of low to moderate LPE conditions. The Si CSR distribution at high LPE values is anomalous and some examples of this are given in the supplementary section. These datapoints were excluded from our PFI analysis.

Using the Si-CSR and Kingham curve, a value of the electric field for each analysis condition was determined. Assuming this electric field value, the corresponding PFI curves for $Si_2$ (Fig. 4(a)), $Si_3$ (Fig. 4(b)) and $Si_4$ (Fig. 4(c)) were constructed. For comparison, the theoretical PFI curves (from Fig. 1) for each cluster size are indicated by the solid lines. The PFI of the Si clusters appear to follow an electric field-dependent trend consistent with the Kingham PFI theory, i.e., the relative abundance of $Si_2^{2+}$ and $Si_3^{2+}$ (blue symbols) clearly increase with electric field. From the experimental data, the electric field value corresponding to $F^{50}$, occurs at ~17.7 V/nm for both $Si_2$ and $Si_3$. Moreover, the $Si_2$ cluster shows a good agreement between the theoretically calculated PFI plot and the experimental data. However, the agreement between theory and experiment is poor for $Si_3$ and $Si_4$. Additionally, for $Si_4$, a crossover point could not be identified from the experimental data alone. One possibility is that we are in an electric field regime greater than $F^{50}$, where the CSR is not as sensitive to the electric field and where the $Si_4^{2+}$ counts have begun to dominate over the $Si_4^{+}$. If this were the case, the data indicates that the crossover point for $Si_4$ would appear to lie ~~anywhere~~ below 17 V/nm (a possible fit to the experimental data is shown in Fig. 5(b)).

Note that for the experimental data, the accuracy of the electric field values (which will effectively correspond to horizontal error bars) is determined by the counting statistics error on the Si-CSR, which when translated to electric field was found to result in an error no larger than ±1%. However, and more importantly, the translation of these Si-CSRs into accurate electric field values is also subject to uncertainty. This is because the Kingham curve of monoatomic Si has not been experimentally validated thus far. Moreover, since there is no direct way to determine the electric field at the specimen apex during an APT experiment, any experimental verification of the Si Kingham curve would only be as accurate as the electric field-calibration method used. Thus, the horizontal error bars will most certainly be much larger. As one would expect, this will also have serious implications for this work, and this will be discussed in the following sections.

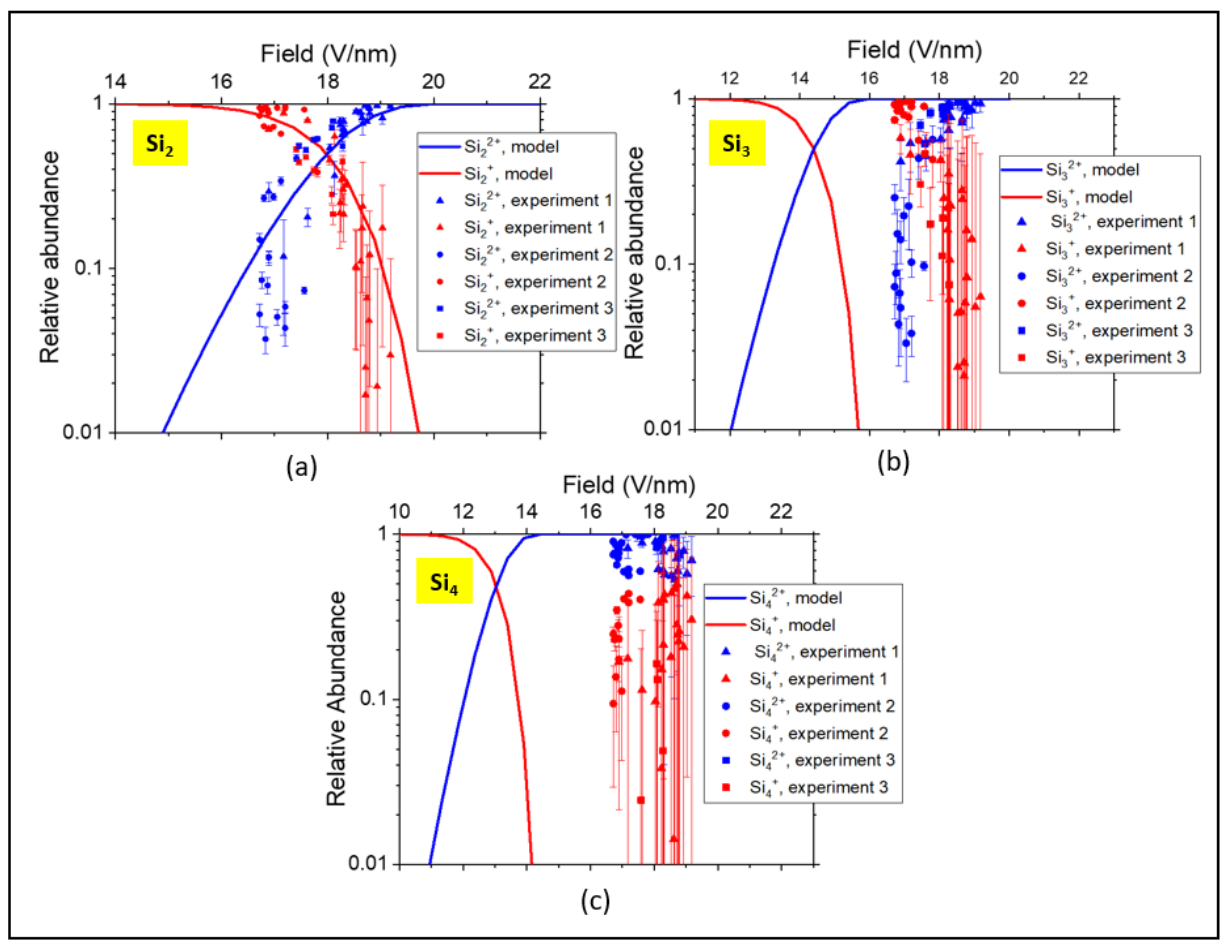

Fig. 4: Scatter plot of the experimental PFI data for (a) $Si_2$, (b) $Si_3$ and (c) $Si_4$. Vertical error bars on experimental data points indicate the $2\sigma$ error due to counting statistics. The solid lines in each figure correspond to the theoretical PFI curve computed using the VIEs calculated by DFT (Table 1).

# 4. Discussion

## 4.1 Comparison of the experimental data with the model predictions

The experimental and theoretical $F^{50}$ values, and the 2$^{nd}$ VIE values calculated by DFT are summarized in Table 2. Both the experimental data and the theory indicate that the PFI of $Si_2$, $Si_3$ and $Si_4$ clusters occurs at a lower electric field compared to Si (re: $F^{50}$ values). However, if we examine the experimental data from Figs 5 (a), (b) and (c) (solid symbols) and compare these, we find that the variation of $F^{50}$ with cluster size does not agree with the model predictions.

| Species | $F^{50}$ – experimental (V/nm) | $F^{50}$ – Theoretical (V/nm) | 2$^{nd}$ VIE (eV) |
|---|---|---|---|
| Si (used as the reference) | 19.6 | 19.6 | 16.4 |
| $Si_2$ | 17.7 | 18.0 | 15.8 |
| $Si_3$ | 17.7 | 14.4 | 14.4 |
| $Si_4$ | < 17.0 | 13.0 | 13.9 |

Table 2: The experimental and theoretical $F^{50}$ values from Figs 5 and 2, respectively. The 2$^{nd}$ VIE computed using DFT (from Table 1) are displayed here to allow a correlation with the $F^{50}$ values. The $F^{50}$ value for Si is taken to be equal to the theoretical value.

To explain the poor fit quality, we assessed the potential impact of the model parameters on the quality of the fit. There are two parameters that require careful examination in the context of clusters, namely, the empirical expression used for the effective nuclear potential, Z, and the ionization energy, I. Additionally, two other model parameters, i.e., the principal quantum number (through the expression for the electron vibration frequency) and the work function, will influence the fit too. However, the sensitivity of the model to these last two quantities is extremely low and it would require unreasonably large changes to these parameters to match the data (details in supplementary section). Thus, in this work, we will limit the discussion to the influence of Z and I on the quality of the fit.

**(1) Effective nuclear potential seen by the tunneling electron:** The Kingham model was a single parameter fit applied to the Rh experimental data. The fitting parameter, Z, is the effective hydrogenic [19] nuclear potential as seen by the tunneling electron and is given by the empirical relation,

$$Z = n + 1 + 4.5/z_0 \qquad \text{E2}$$

where $z_0$ is the distance of the ion from the model surface and n is its initial charge-state. This potential will be a superposition of the electric potential due to the applied electric field and the ionic potential of field-evaporating ion. Thus, Z will essentially be a function of the final ionic charge-state (n +1) as well as the distance $z_0$. For a more detailed explanation on the choice of the above fitting function and its dependence on $z_0$, the reader is referred to [19]. In keeping with Kingham's work, we used a similar function. However, due to the greater screening of the nuclear charge by the electrons in a molecule compared to an atom, it is possible that the effective nuclear charge seen by an electron that is tunneling out from a molecular ion will be lower than what is seen by an electron tunneling from a monoatomic ion. In accordance with this, decreasing the parameter Z has the effect of shifting the PFI curves towards the right, i.e., towards higher electric field, thus resulting in an improved fit to the experimental data. Since the model is a good fit to the $Si_2$ experimental data (Fig. 4 (a)), this suggests that the original empirical function for Z (E2) is still an acceptable approximation for $Si_2$. However, for the larger clusters, it is possible that E2 overestimates Z. To improve the quality of the fit for $Si_3$ and $Si_4$, the coefficients of this polynomial were varied such that the theoretical and experimental $F^{50}$ values agreed. A reasonably close fit to the experimental data was obtained by using the empirical expressions $Z' = n + 0.55 + 1/z_0$ for $Si_3$ and $Z' = n + 0.28 + 1/z_0$ for $Si_4$ (Fig. 5).

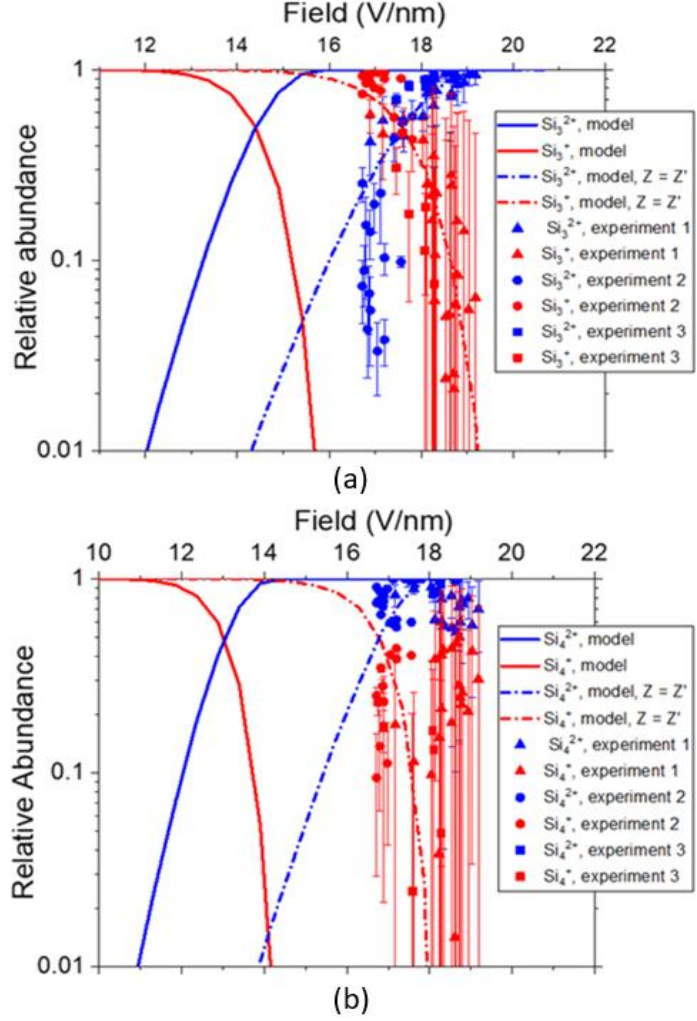

Fig. 5: Impact of decreasing Z on the PFI model (dashed lines) for (a) $Si_3$, using $Z' = n + 0.55 + 1/z_0$ and (b) $Si_4$, using $Z' = n + 0.28 + 1/z_0$. Also shown (solid lines) are the PFI curves computed using Z given by E2.

**(2) Cluster ionization energy:** For any given electric field the PFI probability is extremely sensitive to the ionization energy, I (ES35 in supplementary section). Given that the 2$^{nd}$ VIE of the Si clusters decreases with increasing size, the model predicts that the $F^{50}$ values should also decrease monotonously. However, our experimental results would indicate that $Si_2$ and $Si_3$ have approximately the same experimental $F^{50}$ value (~17.7 V/nm). The accuracy of IEs computed by DFT depends on the approximations used. Thus, it is standard practice to benchmark the approximations (primarily, the energy functional) against experimental data of any other parameter that may be available for the same system. As was explained in section 2.3., the approximations chosen for the Si cluster VIE calculations were benchmarked against experimental data of the 1$^{st}$ AIE of the Si clusters ([40], supplementary section). The intrinsic computational error on the VIE values is therefore not expected to be greater than ~0.2 eV [42]. However, an important effect that is not considered in the DFT calculations presented here is the impact of the electric field on the IEs. In an electric field the molecular orbitals will become polarized, and the electrons will interact with the electric field. This has an impact on how the electrons interact with the ion core and leads to the electronic energy levels becoming more negative [19], effectively increasing the IE of the molecule. The deviation of the model predictions (for $Si_3$ and $Si_4$) from the experimental data could potentially be explained by the effect of the electric field on the IE of these clusters. For these clusters and by varying the VIE values, a better agreement between the theory and experimental data is achieved with a ~9.8% and ~13% increase of the $2^{nd}$ VIE for $Si_3$ and $Si_4$, respectively (dashed lines, Fig. 6 (a) and (b)). This corresponds to an increase of the $2^{nd}$ VIE by 1.41 eV ($Si_3$) and 1.55 eV ($Si_4$). It is unclear whether this magnitude of electric field-induced IE shift is realistically possible and is a non-trivial question to answer. Although there is some recent literature about the impact of the electric field on the molecular orbitals in the context of APT using ab-initio and molecular dynamic simulations [43–45], to the best of the authors' knowledge, there are currently no reported studies on the quantitative impact of the electric field on the IEs of molecular ions. While it would be worthwhile to investigate this further, at this stage we note that it is more practical to treat Z as the only fitting parameter while using electric field-free VIE values in the model, primarily because of the complexity involved in computationally/experimentally estimating the shift of the clusters VIEs due to an electric field.

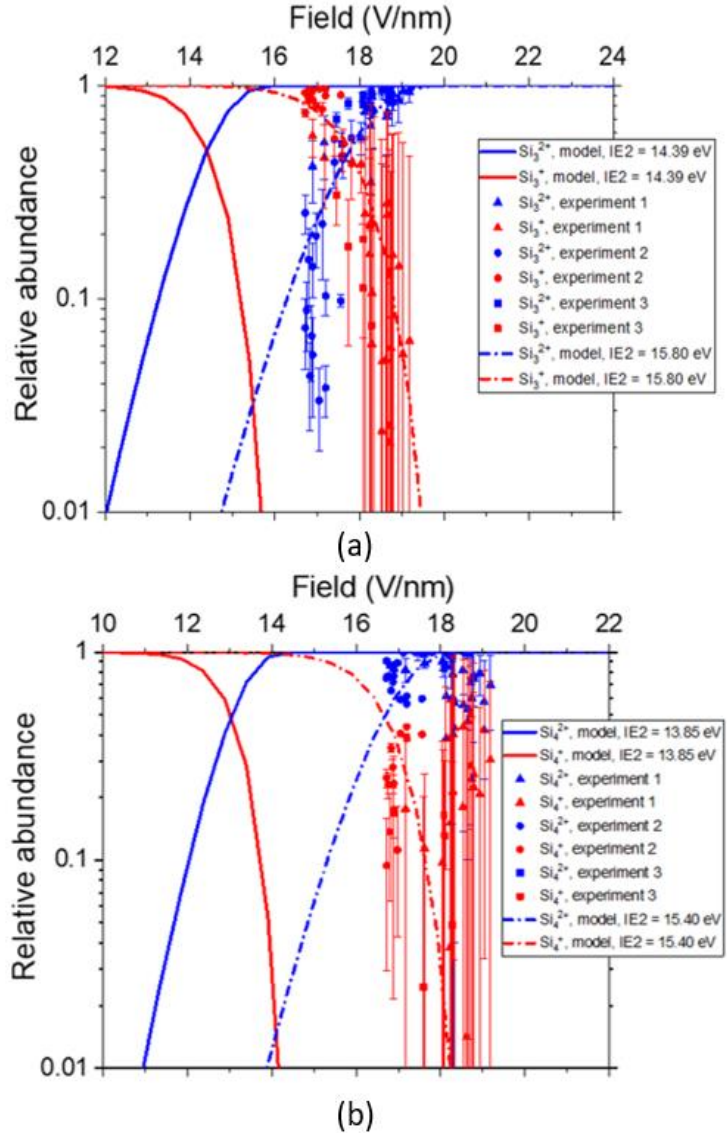

Fig. 6: Impact of varying the 2$^{nd}$ VIE on the PFI model (dashed lines) for (a) Si$_3$ and (b) Si$_4$. Also shown (solid lines) are the PFI curves computed using the electric field-free VIE values (Table 1). In (a), the solid lines were calculated using an IE2 = 14.39 eV, whereas the dashed lines were calculated using an IE2 = 15.80 eV to obtain a better fit to the data. Similarly, in (b), the solid lines were calculated using an IE2 = 13.85 eV, whereas the dashed lines were calculated using an IE2 of 15.40 eV to obtain a better fit to the data.

## 4.2 Resolving peak overlap for monoisotopic species using cluster PFI curves – challenges and ways forward

To assess whether the above conclusions are transferrable to another system and can be used to resolve peak overlap between clusters of monoisotopic species, as proposed initially, we considered the PFI of As clusters in the APT of InGaAs. The electric field-dependence of the InGaAs composition determined by APT was investigated in detail in a previous study [17], which revealed that the As content was underestimated for nearly all operating conditions. From the analysis, it was concluded that mass peak overlaps were the most important source of compositional inaccuracy. From that study, we chose an operational electric field regime (laser pulse energy = 0.1 pJ; In-CSR = 0.0056), where the As content determined by APT was 46.5 at. % (whereas its nominal value should be 50 at. %). The mass spectrum acquired at these conditions is shown in Fig. 7, with all the As cluster peaks indicated. The $As_6^+$ mass peak (450 Da) is absent and, thus, it is reasonable to assume that $As_6^{2+}$ is not emitted. Moreover, the $As_5$ is also absent and the $As_5^{2+}$ peak has negligible counts, which further justifies this assumption. This means that there will be no interference with the $As_3^+$ at 225 Da. The inset in Fig. 7 shows the number of ionic counts at each mass peak. Two of these peaks could potentially be impacted by overlap, namely, the peak at 150 Da, which could contain $As_4^{2+}$ and $As_2^+$, and the peak at 75 Da, which could contain $As_2^{2+}$ and $As^+$. In attempting to resolve these two overlaps using PFI theory, four main inconsistencies in the results were found, which indicated that further refinements to this approach will be required before it can be used as a standalone technique for resolving overlap. The procedure followed and the inconsistencies encountered therein are described in detail below.

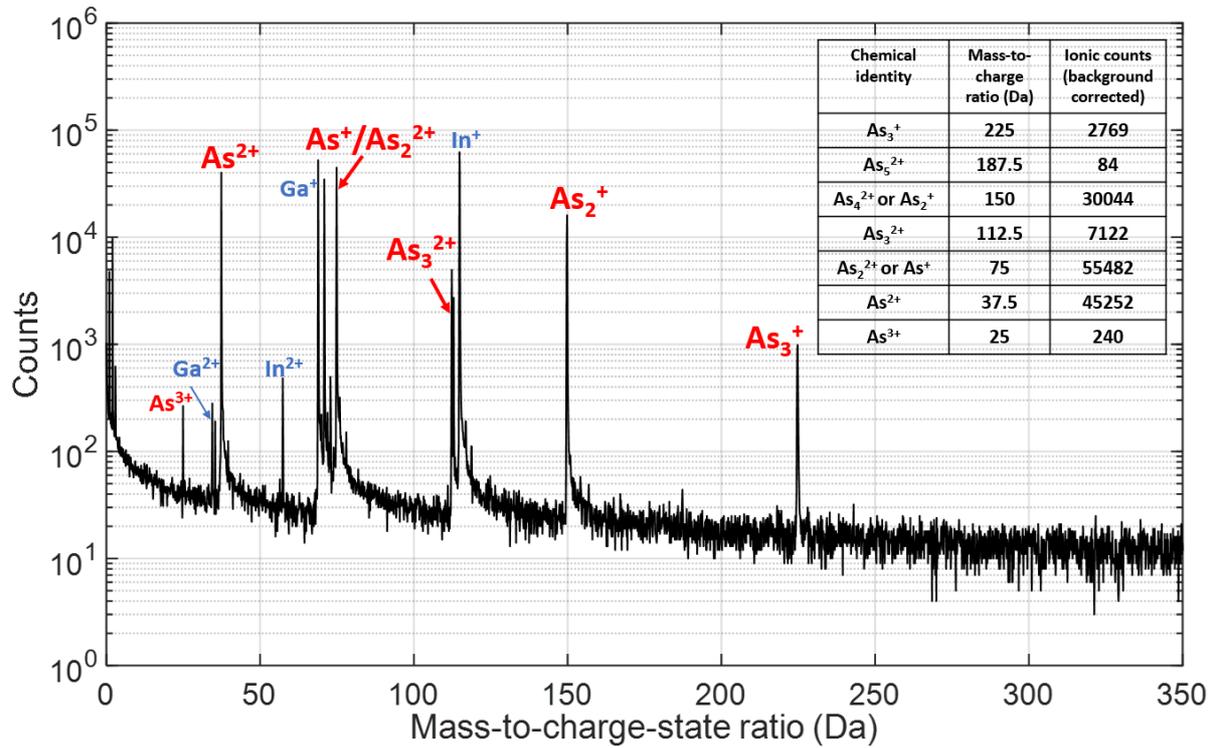

Fig. 7: The mass spectrum of InGaAs acquired on the CAMECA LEAP 5000XR at a laser pulse energy of 0.1 pJ and an In-CSR of 0.0056. The inset indicates the total number of ionic counts present in each peak corresponding to the As clusters

1. The In-CSR together with the In Kingham curve was first used to calibrate the experimental data against the electric field. For this case, the electric field was determined to be 21.3 V/nm.
2. The theoretical PFI curves of $As_2$, $As_3$ and $As_4$ were then computed using the corresponding expression for Z' that best matched the $Si_2$, $Si_3$ and $Si_4$ experimental data respectively. These three As cluster PFI curves along with the monoatomic As PFI curve are indicated in Fig. 8 by the solid lines. From these curves, at an electric field of 21.3 V/nm, we obtain the following.
    (a) $As^{2+}/ (As^{2+} + As^+) = 0.0143$
    (b) $As_2^{2+}/ (As_2^{2+} + As_2^+) = 1.0000$
    (c) $As_3^{2+}/ (As_3^{2+} + As_3^+) = 0.9468$
    (d) $As_4^{2+}/ (As_4^{2+} + As_4^+) = 1.0000$.
3. From 2(b), the counts of $As_2^+$ should be zero. This essentially means that any counts present at 150 Da should belong to $As_4^{2+}$. However, upon assigning all counts at 150 Da to $As_4^{2+}$, we obtain a revised As content of 53 at. %, which exceeds the nominal value and thus is an incorrect prediction.

4. Furthermore, since the As$_3$ and As$_3^{2+}$ peaks in this case do not suffer from any overlaps, it is possible to compare the theoretical As$_3$ PFI curve with experiment (Fig. 8(c), solid lines versus solid symbols). Unfortunately, the agreement with the theoretical predictions is poor. Upon extrapolation, the experimental F$^{50}$ value will most likely be slightly greater than the theoretical value of 19.7 V/nm.

5. At 75 Da, we have 55482 background corrected counts, some of which are As$_2^{2+}$ and the rest are As$^+$. Using 2(a) and the counts at 37.5 Da (As$^{2+}$), we can compute the fraction of counts that should be attributed to As$^+$. However, this leads to unrealistic results, i.e., the model's prediction of the As$^+$ counts exceeds the available counts in the 75 Da peak.

6. Additionally, the presence of As$^{3+}$ at 25 Da in the mass spectrum (Fig. 7) is unexpected since it is inconsistent with the monoatomic As PFI curve (Fig. 8(a)) according to which, As$^{3+}$ should not be observed at an electric field of 21.3 V/nm.

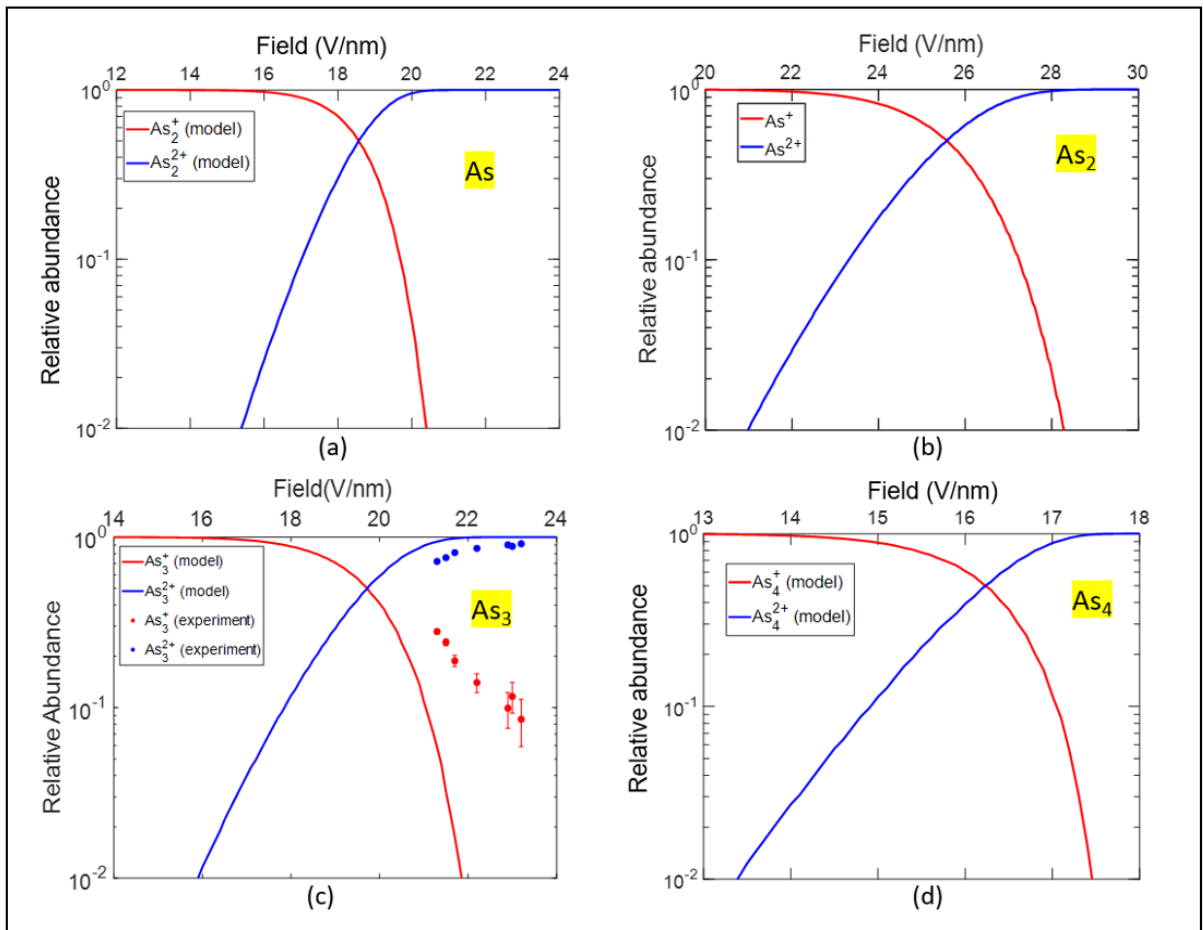

Fig. 8: Theoretical PFI curves (solid lines) of (a) As, (b) As$_2$, (c) As$_3$ and (d) As$_4$. Also indicated in (c) are the experimental As$_3$ data (solid symbols).

There are two bottlenecks for achieving reliable results. Firstly, the PFI model currently does not accurately represent the physical reality, especially for molecules. Since the Kingham model has been verified experimentally for very few elements and the approach proposed in this work relies on the assumption that this model would work equally well for all other elements, it leads to a huge uncertainty in the results. If this assumption were, in fact, true then using any monoatomic CSR (and its PFI curve) to estimate the electric field and subsequently estimate the overlap between clusters at that electric field, would yield accurate results. Since the validity of this assumption cannot be established currently, any electric field determination method based on the Kingham model will be rendered arbitrary. This essentially means that the Kingham model itself would benefit from further refinement, i.e., PFI theory for both molecular and monoatomic ions deserve more attention. Based on the criticism of Kingham's model in later work [46], it is worth examining alternative approaches to calculate the electron tunneling rate constant. Particularly, for molecules, a more complex approach may be required. Moreover, the decay of the electric field strength as a function of the distance from the emitter surface and its impact on the PFI probability also need to be accounted for. Additionally, there is also a need for standardizing the PFI model/equations used and its calculation across the APT community.

Secondly, the development of a more refined PFI model would necessarily require experimental data against which the model can be validated. To obtain such (reliable) experimental data, an accurate electric field determination/calibration procedure would be required, and this presents an additional challenge. As discussed in [47], there are currently no tested methods to experimentally measure (directly or indirectly) the average electric field at the apex of the APT specimen. This is because, the radius of the specimen at any given instant during data acquisition is unknown. Upcoming in-situ techniques that would potentially be able to probe this information involve the incorporation of an atom probe within a transmission electron microscope, which in itself is a massive undertaking and still in an early developmental stage [48].

# 5. Conclusions

This work was the first experimental investigation of the PFI behavior of molecular/cluster ions in APT. We have shown that the charge-state variation of Si clusters $Si_2$ and $Si_3$ as a function of the electric field is qualitatively consistent with the predictions of PFI theory, i.e., the fraction of doubly charged species increases with the electric field. However, the variation of the $F^{50}$ value as a function of the cluster size showed a poor agreement between the experimental data and theoretical predictions. The main reason for the quantitative disagreement is the simplicity of assumptions made in the model parameters when extending the Kingham model, which was originally developed for monoatomic ions, to molecules. Based on experimental data of the Si clusters, the model parameters could be re-calibrated to yield an improved fit. However, upon applying this revised model to resolve peak overlap between As clusters, several inaccuracies were found, thus indicating that PFI theory (for both monoatomic and molecular species) and electric field-calibration methods need further development. If a more detailed model can be developed in future, then it would be worth re-evaluating the transferability of the proposed concept and its feasibility for resolving peak overlap.

## 6. Supplementary Material

The supplementary section of this manuscript contains a detailed description of the Kingham PFI model, our calculations of the PFI curves and the details of the DFT calculations of the Si cluster ionization energies.

## 7. Acknowledgements

R.C., R.J.H.M., C.F. and W.V. acknowledge the financial support by FWO-Hercules through project ZW13_09. P.F. acknowledges the Research Foundation−Flanders (FWO) for a senior postdoctoral grant.

# Supplementary material: Post-field ionization of Si clusters in atom probe tomography: A joint theoretical and experimental study

Ramya Cuduvally*[1,2], Richard J. H. Morris[2], Giel Oosterbos[1], Piero Ferrari[1], Claudia Fleischmann[1,2], Richard G. Forbes[3], Wilfried Vandervorst[1,2]

[1.] *Quantum Solid-State Physics, KU Leuven, Celestijnenlaan 200D, 3001 Leuven, Belgium.*

[2.] *IMEC, Kapeldreef 75, 3001, Leuven, Belgium.*

[3.] *Advanced Technology Institute & Department of Electrical and Electronic Engineering, University of Surrey, Guildford, Surrey GU2 7XH, UK.*

*\*Current address: Canadian Centre for Electron Microscopy, McMaster University, 1280 Main St West, Hamilton, Ontario, L8S 4M1, Canada*

## 1. Theory

### 1.1 Kingham's PFI model

The early experimental observations for the CSR electric field dependence of field-evaporated ions could not be explained by either of the existing image-hump [1] or charge-exchange [2] Field-Evaporation (FEV) models. Today, it is generally accepted that multiply charged ions in APT may be produced upon field-evaporation or PFI. In this section, we will provide a description of the analytical formulation of PFI as presented by Kingham [3] and then attempt to reproduce the Kingham curves for Rh and Si. The purpose here is to elucidate on the aspects which are unclear, along with the challenges faced in reproducing the original work. We will refer to Kingham's paper [3] as Paper 1/Kingham's (PFI) model, going forward.

From Paper 1, the PFI probability, $P_t$, is given by,

$$P_t = 1 - \exp\left[-\int_{z_c}^{\infty} R(z_0) u(z_0)^{-1} dz_0\right] \qquad \text{E(S1)}$$

Here, $z_0$ is the distance from the ion core to the emitter surface, $R(z_0)$ is the ionization rate-constant or equivalently the electron tunneling rate-constant, $u(z_0)$ is the ion velocity as it is being accelerated away from the emitter surface and $z_c$ is the minimum value of $z_0$ beyond which PFI can occur (called the critical distance). In the following sections, we will describe each of these parameters in detail. In the PFI theory, it is assumed that the field-evaporated ion has an initial charge-state of +1, i.e., the relative abundance of the +1 charge-state is taken to be unity until PFI from the +1 to +2 charge-state begins to occur beyond a certain value of field. Note that in Kingham's model, the decay of the electric field strength as a function distance from the emitter was not considered and the analytical treatment is one dimensional.

### 1.1.1 Critical distance, $z_c$

Consider the following scenario. An ion is initially field-evaporated from a positively biased metal emitter in the +1 charge-state. We are now interested in the PFI of this ion from the +1 to the +2 charge-state. This can happen in two ways. (1) The +1 ion may lose an additional electron by tunneling into an available energy state above the emitter Fermi level, and (2), If the outermost electron in the +1 ion aligns with a state above the vacuum level, then the electron may tunnel from the +1 ion into free space. Both these processes, however, can only occur outside/beyond a "crossing surface", which is defined as the space outside which the energy level of the outermost electron of the +1 ion lies above the Fermi level of the emitter. In a 1D treatment, the crossing surface is called the crossing point and it is defined as the point at which the potential energy curves of the +1 and +2 ion intersect. In FEV theory, this distance, which is typically measured from the image plane or electrical surface (as indicated in Fig. 3 in Paper 1), is called the critical distance, $L_c$. The image plane is that which coincides with the center of gravity of the charge distribution induced by the presence of the +1 ion near the emitter surface. If electric field penetration is disregarded, one can define a model surface as indicated in Fig. 3 in Paper 1, which can alternatively be used as a reference plane (instead of the image-plane) from which distances are measured. In Kingham's PFI model, the critical distance is measured from this model surface and is indicated by $z_c$. Thus, $z_c$ and $L_c$ are simply related by $z_c = L_c - \lambda$, where $\lambda$ is the electric field screening length. Note that $\lambda$ is small and could, in principle, be neglected. In Kingham's PFI model, the model surface is the reference plane from which all distances are measured and the variable $z_0$ is used to indicate the distance of the ion core from the model surface.

To derive the expression for $L_c$ (and subsequently $z_c$), it is necessary to first define the potential energy of the +1 and +2 ion as a function of distance from the image plane. This is typically defined in terms of the standard system potential energies [4], $U_n(L)$, where $L$ is the distance of the ion core from the image plane (Fig. 3, Paper 1) and n is its charge-state. For +1 to +2 charge-state PFI of an ion, $L_c$ will correspond to the value of $L$ for which,

$$U_1(L) = U_2(L) \qquad \text{E(S2)}$$

To define $U_n(L)$ for a position $L$, that can be close to the emitter surface, we start with a neutral atom located within a remote field-free space and take this as the energy zero for all atom/ion potential energies $U_n$. We then consider a formal process that mimics the reality of the situation. In this process, we remove $n$ electrons from the neutral atom one by one and sequentially return them to the metal. According to the details given in [5], this leads to four terms which in their simplest approximation are as follows:

(1) The energy required to create an ion of charge $ne$ in remote field-free space. This is the sum of the first $n$ free-space ionization energies $I_n$.
(2) The energy gained in moving $n$ electrons from $L$ into the emitter and placing them at the Fermi level. In terms of the emitter local work function $\phi$, this is $-n\phi$.
(3) The energy gained in returning the ion (of charge $ne$) from a remote field-free space to L. This gives rise to the image potential energy ($E_{image}$) term, which accounts for the potential due to the image-charge induced by the presence of the ion near a surface. The image force is attractive in nature and has the effect of reducing the critical distance. The image potential energy is given by [4],

$$E_{image} = -\frac{n^2 e^2}{16\pi\varepsilon_0 L} \qquad \text{E(S3)}$$

This can be conveniently written in terms of the image-potential energy constant, C as,

$$E_{image} = -\frac{n^2 C}{L} \qquad \text{E(S4)}$$

where, $C = \frac{e^2}{16\pi\varepsilon_0}$.

(4) The "applied field" component of the energy gained in moving $n$ electrons from $L$ into the emitter. In terms of the local electrostatic field, $F$, this is given by,

$$E_{electrostatic} = -neFL \qquad \text{E(S5)}$$

Additionally, any shift in the Ionization Energy (IE) due to the applied electric field can also be included by an additional term (referred to as the Stark shift of the IEs, $E_{Stark}$). This term can be ignored in a first approximation.

Therefore, $U_1(L)$, $U_2(L)$, $U_3(L)$ and $U_4(L)$ can be written in the International System of Quantities (ISQ)/SI units as:

$$U_1(L) = I_1 - \phi - eFL - C/L \qquad \text{E(S6)}$$

$$U_2(L) = I_1 + I_2 - 2\phi - 2eFL - 4C/L \qquad \text{E(S7)}$$

$$U_3(L) = I_1 + I_2 + I_3 - 3\phi - 3eFL - 9C/L \qquad \text{E(S8)}$$

$$U_4(L) = I_1 + I_2 + I_3 + I_4 - 4\phi - 4eFL - 16C/L \qquad \text{E(S9)}$$

Equating E(S6) and E(S7) gives,

$$eFL_c = I_2 - \phi - 3C/L_c \qquad \text{E(S10)}$$

where the last term is referred to as $\Delta E_{image}$. Rewriting the above equation in terms of $z_c$, we have,

$$eF(z_c + \lambda) = I_2 - \phi - \Delta E_{image}$$

or

$$eFz_c = I_2 - \phi - \Delta E_{image} - eF\lambda \qquad \text{E(S11)}$$

The formulation given in Paper 1 is in the Hartree system of units. In this system, e = $4\pi\varepsilon_0$ = $m_e$ = $\hbar$ =1 and the unit of energy is the Hartree (1 Hartree ~ 27.2 eV). Ignoring the stark shift ($\Delta E_{stark}$) of the IEs, equation E(S11) is equivalent to the expression for $z_c$ given in Paper 1 and was used to compute the Kingham curves in this study, i.e.,

$$Fz_c = I_{n+1} - \Delta E_{stark} - \Delta E_{image} - \phi - \lambda F \qquad \text{E(S12)}$$

Solving the quadratic equation E(S10) we obtain the following expression for the critical distance in ISQ/SI units for the PFI from +1 to +2 charge-state as,

$$L_{c12} = [(I_2 - \phi) \pm \{(I_2 - \phi)^2 - 3eFW\}^{1/2}]/2eF \qquad \text{E(S13)}$$

Similarly, for the PFI from +2 to +3, we have,

$$L_{c23} = [(I_3 - \phi) \pm \{(I_3 - \phi)^2 - 5eFW\}^{1/2}]/2eF \qquad \text{E(S14)}$$

And from +3 to +4,

$$L_{c34} = [(I_4 - \phi) \pm \{(I_4 - \phi)^2 - 7eFW\}^{1/2}]/2eF \qquad \text{E(S15)}$$

Here,

$$C = e^2/16\pi\varepsilon_0 = 0.3599911 \text{ eV nm} \qquad \text{E(S16)}$$

and,

$$W = e^2/4\pi\varepsilon_0 = 4C = 1.439965 \text{ eV nm.} \qquad \text{E(S17)}$$

As an example, for the PFI of $Rh^+$ to $Rh^{2+}$ at 25 V/nm (the field value at which the experimental data by Ernst [6] indicates that the concentration of $Rh^+$ and $Rh^{2+}$ are equal) and with $I_2$ = 18.08 eV and $\phi$ = 4.8 eV, we obtain a value of 0.43 nm for $L_{c12}$, using E(S13).

Note that Paper 1 only lists E(S12) and the above equations, i.e., E(S13), E(S14) and E(S15) while their origin are not explicitly given. Moreover, since all other equations given in Paper 1 are in Hartree units, for clarity, the Hartree-equivalent of equations E(S13), E(S14) and E(S15) are given below,

$$L_{c12} = [(I_2 - \phi) \pm \{(I_2 - \phi)^2 - 3F\}^{1/2}]/2F \qquad \text{E(S18)}$$

$$L_{c23} = [(I_3 - \phi) \pm \{(I_3 - \phi)^2 - 5F\}^{1/2}]/2F \qquad \text{E(S19)}$$

$$L_{c34} = [(I_4 - \phi) \pm \{(I_4 - \phi)^2 - 7F\}^{1/2}]/2F \qquad \text{E(S20)}$$

### 1.1.2 Kinetic energy (k) and ion velocity (u)

The expression for the kinetic energy of the field-evaporated ion used in Kingham's model (in Hartree units) is given by,

$$\frac{1}{2}mu(z_0 + \lambda)^2 = nF(z_0 + \lambda) - \sum_{r=n_i}^{n-1} F(z_r + \lambda) - \sum_{r=n_i}^{n-1} \frac{(2r+1)}{4(z_r + \lambda)} + \frac{n^2}{4(z_0 + \lambda)} - (n_i^3 F)^{1/2}$$

$$\text{E(S21)}$$

Here, m is the atomic mass, u is the velocity and F is the electric field strength. The variable $z_r$ corresponds to the critical distance where the PFI from charge-state r to r +1 occurs. The variables n and $n_i$ represent the charge-state but the distinction between them (in the first and last terms) is unclear from Paper 1. Since the derivation/origin of the expression is not given in the original work, we attempt to derive it here in ISQ/SI units first, and then compare our expression with E(S21). We will consider the PFI of a field-evaporated ion in its initial +1 charge-state and its PFI transition into the +2 charge-state. It is assumed that after its escape, the 1+ ion travels in a potential energy that can be described by the quantity $U_1(L)$ as given in E(S6). From E(S17), we have, C = W/4. Thus, E(S6) becomes,

$$U_1(L) = I_1 - \phi - eFL - W/(4L) \qquad \text{E(S22)}$$

We assume that the +1 ion escapes/field-evaporates at the top of a Schottky image hump located at $L = L_i$. At this point, the ion has a potential energy of $U_1(L_i)$ and zero kinetic energy, k. At any given position L ($L > L_i$), the ion will have picked up kinetic energy, k(L) given by the difference in potential energy, i.e.,

$$k(L) = U_1(L_i) - U_1(L) \qquad \text{E(S23)}$$

Thus, at L = Lc, we have,

$$k(L_c) = U_1(L_i) - U_1(L_c) \qquad \text{E(S24)}$$

To determine $U_1(L_i)$, we differentiate E(S22) with respect to L and equate it to zero, to find the L-value corresponding to the maxima, i.e.,

$$dU_1/dL = -eF + W/(4L^2) = 0 \qquad \text{E(S25)}$$

From the above equation and at L= $L_i$ we have,

$$W/4L_i = eFL_i \qquad \text{E(S26)}$$

and substituting this into E(S22) yields,

$$U_1(L_i) = (I_1 - \phi) - 2eFL_i \qquad \text{E(S27)}$$

It can also be deduced from E(S26) that,

$$L_i = (1/2)(W/eF)^{1/2} \qquad \text{E(S28)}$$

And substituting $L_i$ into E(S22) yields,

$$U_1(L_i) = (I_1 - \phi) - (WeF)^{1/2} \qquad \text{E(S29)}$$

Here we assign, $\qquad (We)^{1/2} = (e^3/4\pi\varepsilon_0)^{1/2} = c_S = 1.199985$ eV (V/nm)$^{-1/2}$

This leads to,
$$U_1(L_i) = (I_1 - \phi) - c_s F^{1/2} \qquad \text{E(S30)}$$

As given in the previous section, $U_1(L_c)$ can be written as,

$$U_1(L_c) = (I_1 - \phi) - eFL_c - W/(4L_c) \qquad \text{E(S31)}$$

Thus, substituting E(S29) and E(S30) into E(S24) yields,

$$k(L_c) = U_1(L_i) - U_1(L_c) = eFL_c + W/(4L_c) - c_s F^{1/2} \qquad \text{E(S32)}$$

The above equation is in the ISQ/SI unit system of equations. Thus, in the Hartree-unit system, it becomes,

$$k_H(L_c) = F_H L_{cH} + 1/(4L_{cH}) - F_H^{1/2} \qquad \text{E(S33)}$$

The above equation is physically identical to E(S21) upon setting n=n$_i$=1 for the PFI from +1 to +2. This also implies that both the summation terms vanish, and so,

$$\frac{1}{2}mu(z_0 + \lambda)^2 = nF(z_0 + \lambda) + \frac{n^2}{4(z_0 + \lambda)} - (n_i^3 F)^{1/2} \qquad \text{E(S34)}$$

The equations E(S33) and E(S34) are both written in Hartree-units and are equivalent, and thus validating our interpretation and derivation of ion velocity used in Paper 1. Based on this understanding and using E(S32) (ISQ) or E(S33)/E(S34) (Hartree), for Rh$^+$ to Rh$^{2+}$ at 25 V/nm and L= L$_c$ = 0.43 nm, the kinetic energy k, is determined to be 5.58 eV.

The same approach for the PFI of +2 to +3 and +3 to +4 would have to be carried out in detail but since these cases are not critical for this work, it will not be covered here. Although irrelevant for the PFI from the +1 to the +2 charge-state, the details of the origin of the two summation terms given in E(S21) can be found in [7].

### 1.1.3 Ionization rate constant/Electron tunneling rate-constant, R

The calculation of the ionization rate constant or the electron tunneling rate-constant, R, is the most influential parameter for the computation of the PFI curves. It entails solving the (time-independent) Schrödinger equation for an electron under the influence of an ionic and electric field potential. The electron is assumed to tunnel through a rounded triangular barrier. The electron wave-function is defined when it is inside the barrier, outside the barrier and then inside the metal/emitter. The wave-function inside the barrier is then matched to a JWKB-type solution [8]. Finally, the total emitted electron flux is found by integrating the emitted flux density (flux per unit area) across a plane (inside the metal) parallel to the metal-model surface. These results are then compared with the Landau and Lifshitz theory [9], which is the most widely accepted theory of field-ionization. A normalization factor/numerical pre-factor (A$^2$ν) is deduced which has the effect of minimizing discrepancies in their theoretical derivation. All the details of this derivation and the mathematical approximations used therein can be found in Paper 1 which yielded the following final expression for R($z_0$).

$$R(z_0) = \frac{6\pi A^2 \nu F}{2^{5/2}[I^{3/2} - (I - ZF/I - Fz_0)^{3/2}]} \left(\frac{16I^2}{ZF}\right)^{\sqrt{Z(2/I)}}$$
$$\times exp[-2^{5/2}I^{3/2}/3F + Z(2/I)^{1/2}/3 + 2^{5/2}(I - ZF/I - Fz_0)^{3/2}/3F]$$
E(S35)

Here, I is the (n +1)$^{th}$ ionization energy for any given charge-state, n, Z is the effective nuclear potential as seen by the tunneling electron, A is a constant, and ν is the electron vibrational frequency. $A^2\nu$ is given by [3],

$$A^2\nu = \frac{I_{n+1}}{6\pi . m_q . \exp(2/3)}$$
E(S36)

where $m_q$ is the principal quantum number of the outermost electron and $A^2$ is found by imposing the requirement that the ionization rate constant for a hydrogen atom in uniform field yields the correct form of R [3]. R has the units of s$^{-1}$ in ISQ. In atomic units, the unit of time is much smaller, whereby 1 [s$^{-1}$] = 2.418884326509 × 10$^{-17}$ [a.u.].

## 1.2 Review of PFI literature

The criticism of the theoretical approach employed in Paper 1 is also worth examining. Lam and Needs [10] pointed out a misprint in E(S35) thus having to replace the factor of $\sqrt{Z(2/B)}$ by $Z\sqrt{(2/B)}$ in the exponent. They also highlighted other discrepancies in the formula for R and presented a revised expression for the ionization rate constant. Although the authors claimed that their method was more accurate than Kingham's, they did not demonstrate the agreement of their computed PFI probability with any experiments. Our attempts thus far to calculate the PFI probability using the revised expression for R and $A^2\nu$ given by Lam and Needs did not yield a satisfactory fit to the Rh experimental data, and these details are shown in Fig. S1. The integration for $P_t$ is carried from $z_c$ to 100 a.u. Also shown is the evolution of R as a function of $z_0$ and F (Fig. S2).

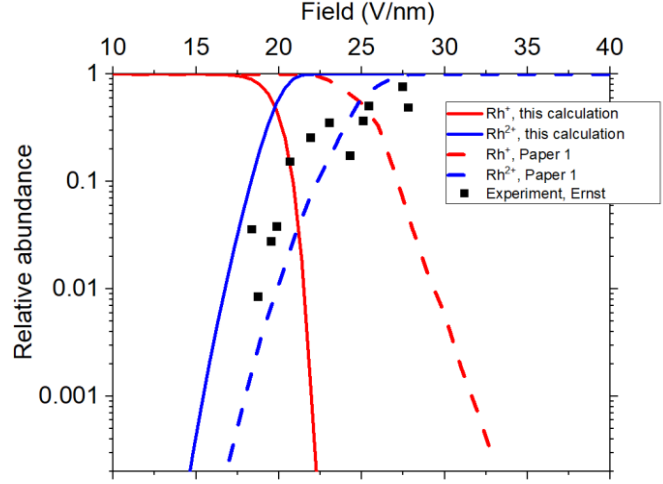

Fig. S1: Comparison of experimental PFI data (symbols) with theoretical curves reported in Paper 1 (dotted lines) and computed in this work after incorporating the corrections and expressions given by Lam and Needs [10] (solid lines).

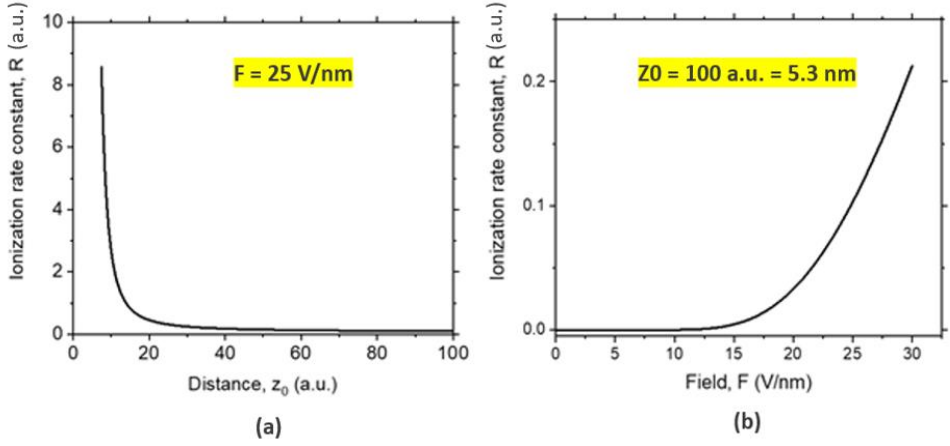

Fig. S2: Ionization rate constant as a function of (a) $z_0$ and (b) F for the PFI of $Rh^+$ to $Rh^{2+}$ computed after incorporating the corrections and expressions given by Lam and Needs [10] (solid lines).

Therefore, in this work, we used E(S37) given below (which is the same as E(S35) but having corrected the misprint in the exponent).

$$R(z_0) = \frac{6\pi A^2 \nu F}{2^{5/2}[I^{3/2} - (I - ZF/I - Fz_0)^{3/2}]} \left(\frac{16I^2}{ZF}\right)^{Z\sqrt{(2/I)}}$$
$$\times exp\left[-2^{5/2}B^{3/2}/3F + Z(2/I)^{1/2}/3 + 2^{5/2}(I - ZF/I - Fz_0)^{3/2}/3F\right] \quad \text{E(S37)}$$

Fig. S3 shows the evolution of R as a function of $z_0$ (a) for F = 25 V/nm and R as a function of the field, F for $z_0$ = 100 a.u., computed using E(S37) given above for the PFI of $Rh^+$ to $Rh^{2+}$.

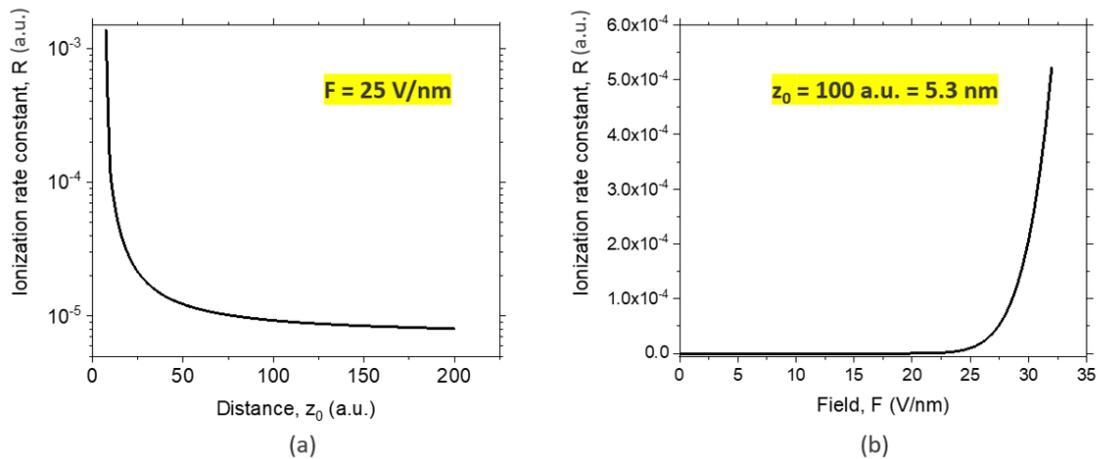

Fig. S3: Ionization rate constant as a function of (a) $z_0$ and (b) F for the PFI of $Rh^+$ to $Rh^{2+}$.

Using the above expression for $R(z_0)$ (E(S37)), we obtain a reasonable agreement with the Rh experimental data (Fig. S4(a)) by integrating $P_t$ (E(S1)) close to the tip surface, i.e., from $z_c$ to 200 a.u. Using the same limits of integration, the curves that we computed for Si are shown in Fig. S4(b). Also shown are the original Kingham curves reproduced from Paper 1. It is unclear whether the misprint referred to by Lam and Needs only appears in the paper or was also carried over in Kingham's computations. Moreover, we are also unsure whether Kingham used the expression for R (E(S35)), which is based on mathematical approximations, or performed a numerical integration of their initial expression (E3.6, Paper 1). We believe that the discrepancy in the shape of the curves (solid versus dotted lines) seen in Figs S5(a) and S5(b) has more to do with the difference in the integration scheme used rather than the misprint. Nevertheless, to study/model the PFI behavior of Si clusters in this work, we believe that our calculations using E(S35) as shown in Fig. S4 are suitable. To model the cluster PFI, we have used E(S33), E(S37) and E(S1) to compute the ion velocity, ionization rate constant and post-field-ionization probability, respectively.

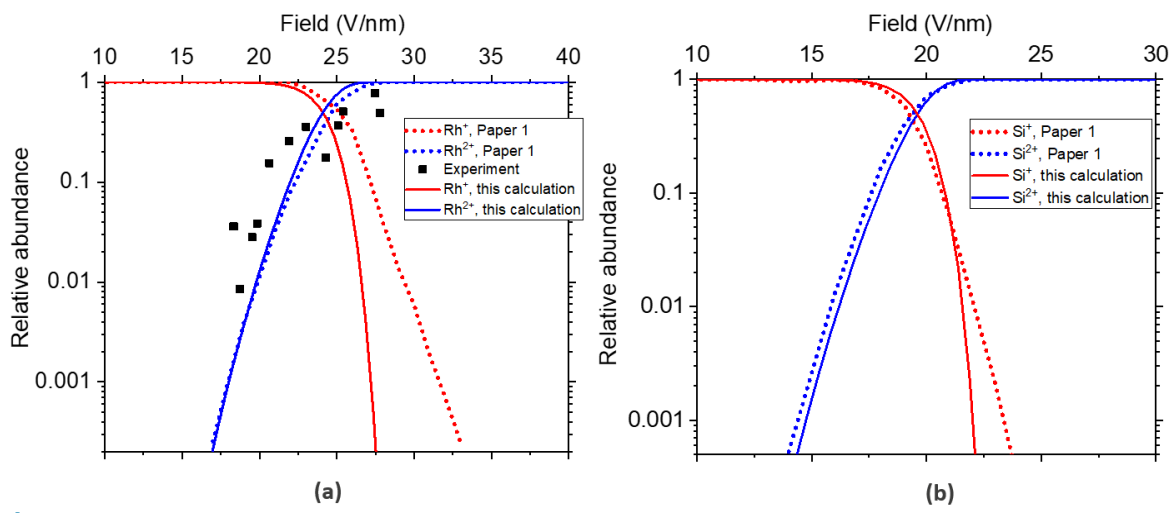

Fig. S4: Kingham curves for Rh (a) and Si (b) computed in this work (solid lines). E(S33), E(S37) and E(S1) were used for computing the ion velocity, ionization rate constant and post-field-ionization probability, respectively. The dotted lines are replicated from Paper 1 using graphreader.com.

## 2. DFT calculations of Si clusters

### 2.1.1 Energy optimized Si Cluster XYZ atomic coordinates(Å)

#### 2.1.1.1 Neutral clusters
$Si_2$:
Si 0.041810236026 0.000000000000 0.000000000000
Si 2.268046763974 0.000000000000 0.000000000000

$Si_3$:
Si -0.057343762281 -0.033200284014 0.000000000000
Si 2.227316799634 -0.033247694419 0.000000000000
Si 1.085026962645 1.945447978435 0.000000000000

$Si_4$:
Si 0.317099540958 0.315074980037 0.000098806249
Si 2.537031740828 -0.233656414502 -0.000097761127
Si 1.982900556508 1.984925114390 0.000095855695
Si -0.237031838294 2.533656320075 -0.000096900815

#### 2.1.1.2 Singly charge clusters
$Si_2^+$:
Si 0.029279961385 0.000000000000 0.000000000000
Si 2.280577038615 0.000000000000 0.000000000000

Si$_3^+$:
Si -0.057343762281 -0.033200284014 0.000000000000
Si 2.227316799634 -0.033247694419 0.000000000000
Si 1.085026962645 1.945447978435 0.000000000000

Si$_4^+$:
Si 0.232974380191 0.230730546789 -0.000183714836
Si 2.463582356761 -0.160395512577 0.000184934441
Si 2.067025167246 2.069270045586 -0.000186962137
Si -0.163581904199 2.460394920202 0.000185742533

### 2.1.1.3 Doubly charged clusters

Si$_2^{2+}$:
Si -0.279162333958 0.000000000000 0.000000000000
Si 2.589019333958 0.000000000000 0.000000000000

Si$_3^{2+}$:
Si -0.202483118101 0.009834378735 0.000000000000
Si 2.372483122005 0.009834377797 0.000000000000
Si 1.084999996095 1.859331243471 0.000000000000

Si$_4^{2+}$:
Si 0.019437474412 0.016679588424 0.000002942630
Si 2.283310546518 0.019447243196 -0.000001291591
Si 2.280562526914 2.283320411514 -0.000001062385
Si 0.016689452156 2.280552756867 -0.000000588652

### 2.1.2 Adiabatic Ionization Energies (AIEs) of the Si clusters (eV) computed by DFT.

The AIEs were calculated using five different functionals and compared to experimental values. From the total average error computed, we conclude that the ωB97X-D3 functional results in the best fit to experiment.

| Cluster | Experiment [11], [12] | TPSS | B3LYP | PBE | ωB97X-D3 | CAM-B3LYP-D3 |
|---|---|---|---|---|---|---|
| Si | 8.15168 ± 0.00003 | 8.21 | 8.03 | 8.20 | 8.10 | 8.11 |
| Si2 | 7.92 ± 0.05 | 7.83 | 7.77 | 7.77 | 7.93 | 7.91 |
| Si3 | 8.12 ± 0.05 | 8.06 | 7.95 | 8.07 | 8.22 | 8.05 |
| Si4 | 8.2 ± 0.1 | 7.71 | 7.68 | 7.70 | 8.04 | 7.91 |
| Si5 | 7.96 ± 0.07 | 7.98 | 7.83 | 7.96 | 7.99 | 7.92 |
| Si6 | 7.8 ± 0.1 | 7.50 | 7.40 | 7.48 | 7.79 | 7.58 |
| Si7 | 7.8 ± 0.1 | 7.73 | 7.60 | 7.71 | 7.98 | 7.77 |

Table S1: First AIE of Si clusters computed by different DFT functionals and compared to experimental values.

## 3. Experimental

### 3.1 2D distribution of Si charge-state at high laser pulse energy during APT

Examples of anomalous CSR distribution at high LPEs for Si are shown in Figs S6 and S7. Note that the laser side (indicated by the arrow) has a higher concentration of $Si^{2+}$, which is contrary to expectation.

(A) 120 pJ

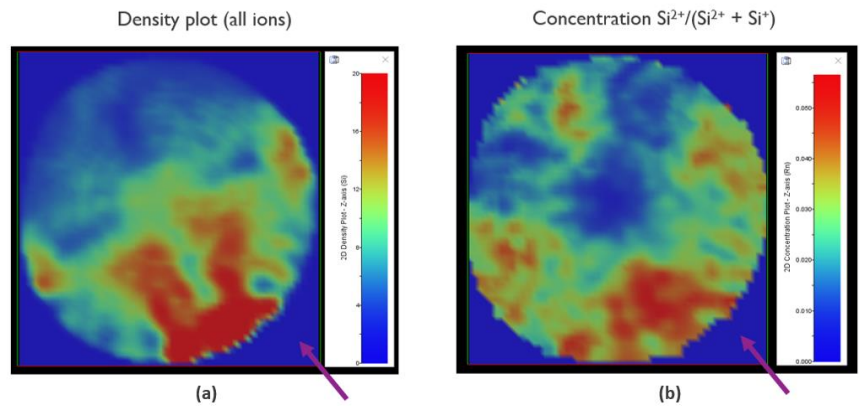

Fig. S5: (a) 2D density plot of all ions and (b) Concentration of $Si^{2+}$ ions at a laser pulse energy of 120 pJ and a detection rate of 1.6 %. The arrow indicates the direction of the UV laser impinging on the specimen.

(B) 155 pJ

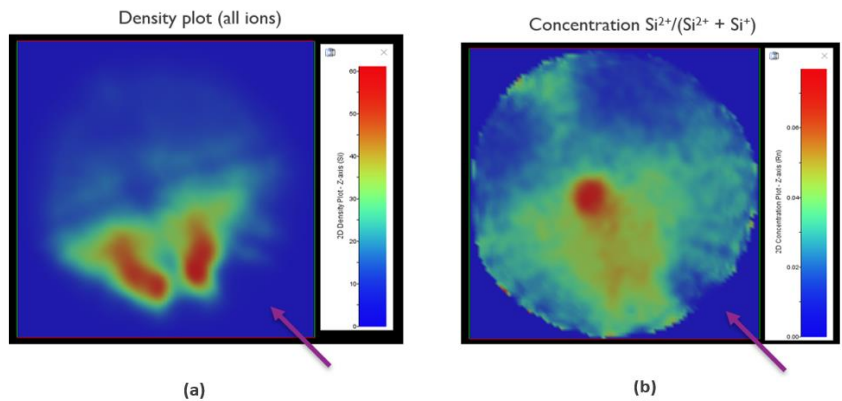

Fig. S6: (a) 2D density plot of all ions and (b) Concentration of $Si^{2+}$ ions at a laser pulse energy of 155 pJ and a detection rate of 1 %. The arrow indicates the direction of the UV laser impinging on the specimen.

## 3.2 Sensitivity of the PFI model to the principal quantum number and surface work function

The principal quantum number, m, feeds into the constant A2v in the expression for R as given in E(S37). It is not straightforward to assign a molecular equivalent to the principal quantum number, which strictly speaking, is an atomic property. For the Si clusters, this was assumed to remain unchanged and is set to 3 (same as Si). Increasing the value of m would decrease R, therefore shifting the PFI curves towards higher field. We found that the PFI curves are not as sensitive to changes in m. For example, for $Si_3$, the experimental and theoretical $F^{50}$ values are 17.7 V/nm and 14.4 V/nm, respectively. However, increasing m by a factor of 3, which is huge, merely increases the theoretical $F^{50}$ from 14.4 V/nm to 15.1 V/nm indicating that the PFI model is not very sensitive to this parameter.

The work function, $\phi$, which is a surface property of the material and is defined as the energy required to remove an electron from the surface Fermi level to a point in vacuum. Several factors such as the crystallographic orientation of the specimen, the presence of dangling bonds, surface dipoles (and their direction) and contaminant gases can affect the value of the work function [3], [13], [14]. Especially on the surface of a positively charged Si field emitter, on which clusters of various sizes are present, the work function may vary locally depending on the cluster size/surface morphology and thus the local charge distribution. A change in the work function would affect the location of the crossing surface and thus the PFI probability, i.e., decreasing the work function would increase the critical distance and decrease the PFI probability since the electron tunneling probability, R, is greater at distances closer to the tip surface. This would again have the effect of shifting the Kingham curves towards higher field. However, for $Si_3$, we found that a 20 % decrease of the work function, which we believe to be significant, increases the $F^{50}$ value from 14.4 to 15.6 V/nm. Thus, the impact, if any, of m and $\phi$ on the quality of the fit is concluded to be a relatively small effect in comparison to the VIE and Z.